\documentclass[aps,preprint,superscriptaddress]{revtex4-1}
\usepackage{placeins}
\usepackage{lmodern} 
\usepackage{float}
\usepackage{amsmath}
\usepackage{amssymb}
\usepackage{amsthm}
\usepackage{amsfonts}
\usepackage{multirow}
\usepackage{graphicx}
\usepackage{hyperref}
\usepackage{graphicx}
\usepackage{epstopdf}
\begin{document}
	%\preprint{}
	
	\title{Resolving ECRH deposition broadening
		due to edge turbulence in DIII-D by heat deposition measurement} 
	\thanks{To be submitted to nuclear fusion}
	
	
	
	\affiliation{Institute for Fusion Studies, Department of Physics, University of Texas at Austin, 2515 Speedway, Austin TX, 78712}
	\affiliation{York Plasma Institute, Department of Physics, University of York, York Y0105DD, UK}	
	\affiliation{General Atomics, 3550 General Atomics Ct., San Diego, CA, 92121}
	
	\author{M W Brookman}
	\email{brookmanmw@fusion.gat.com}
	\affiliation{Institute for Fusion Studies, Department of Physics, University of Texas at Austin, 2515 Speedway, Austin TX, 78712}
	\author{M B Thomas}	
	\affiliation{York Plasma Institute, Department of Physics, University of York, Heslington, York, YO10 5DD UK}
	
	\author{J Leddy}
	\affiliation{York Plasma Institute, Department of Physics, University of York, Heslington, York, YO10 5DD UK}

	\author{C C Petty}	
	\author{R J La Haye}
	\affiliation{General Atomics, 3550 General Atomics Ct., San Diego, CA, 92121}

	\author{K Barada}
	\author{T L Rhodes}
	\affiliation{University of California, Los Angeles, Los Angeles, CA, 90005}
	
	\author{Z Yan}
	\affiliation{University of Wisconsin-Madison, Madison, WI, 53706}

	\author{M E Austin}
	\affiliation{Institute for Fusion Studies, Department of Physics, University of Texas at Austin, 2515 Speedway, Austin TX, 78712}
	
	\author{R G L Vann}
	\affiliation{York Plasma Institute, Department of Physics, University of York, York Y0105DD, UK}

	
	\date{\today}
	
	\begin{abstract}
Interaction between microwave power, used for local heating and mode control, and density fluctuations can produce a broadening of the injected beam, as confirmed in experiment and simulation. Increased power deposition width could impact suppression of tearing mode structures on ITER\cite{refPolipreq}. This work discusses the experimental portion of an effort to understand scattering of injected microwaves by turbulence on the DIII-D tokamak. The corresponding theoretical modeling work can be found in M.B. Thomas \textit{et. al.}:\ \textit{Submitted to Nuclear Fusion} (2017)[Author Note - this paper to be published in same journal]. In a set of perturbative heat transport experiments, tokamak edge millimeter-scale fluctuation levels and microwave heat deposition are measured simultaneously. Beam broadening is separated from heat transport through fitting of modulated fluxes\cite{brookmanec19}. Electron temperature measurements from a 500 kHz, 48-channel ECE radiometer are Fourier analyzed and used to calculate a deposition-dependent flux\cite{stockdale}. Consistency of this flux with a transport model is evaluated. A diffusive($\propto\nabla\tilde{T}_e$) and convective($\propto\tilde{T}_e$) transport solution is linearized and compared with energy conservation-derived fluxes. Comparison between these two forms of heat flux is used to evaluate the quality of ECRF deposition profiles, and a $\chi^2$ minimization finds a significant broadening of 1D equilibrium ray tracing calculations from the benchmarked TORAY-GA ray tracing code\cite{refPrat} is needed. The physical basis, cross-validation, and application of the heat flux method is presented. The method is applied to a range of DIII-D discharges and finds a broadening factor of the deposition profile width which scales linearly with edge density fluctuation level. These experimental results are found to be consistent with the full-wave beam broadening measured by the 3D full wave simulations in the same discharges\cite{thomas}.
	\end{abstract}
	\pacs{4}% insert suggested PACS numbers in braces on next line
	
	\maketitle %\maketitle must follow title, authors, abstract and \pacs
	
	% References should be done using the \cite and \label commands
	
	\section{Introduction}
	 Electron cyclotron current drive (ECCD) is used to stabilize the growth of tearing modes in the plasma through local current drive at the island O-point\cite{lahayeRF}. Past efforts have found agreement in deposition location of the injected microwave beam between experiment\cite{refZer} and ray tracing code TORAY-GA\cite{refPrat} by assuming a level of fast electron transport. However, soft x-ray profile measurements of ECCD on TCV found that fast electrons played a minimal role\cite{refDec}. Rather than being transported from the predicted deposition region, fast electron bremsstrahlung appeared promptly in a wider region. 
	
	  Simulations\cite{koehn16} and analytic work\cite{kyr} have found that density fluctuations of a similar scale to the injected wavelength produce a scattering of the wave, spreading and deflecting the microwave beam. An increase of the current drive width as compared to the tearing mode island width can increase the power requirements of mode suppression. Increased auxiliary heating requirements could impact the fusion gain on ITER if continuous suppression is required\cite{lahayeRF}.
	 
	 That concern motivates this effort on the DIII-D tokamak to understand fluctuation broadening in tokamak discharges through measurement of microwave power deposition. This work will introduce a means of transport fitting to measure the deposition profile width, which scales with the measured level of millimeter-scale density fluctuations in the tokamak edge. An order of magnitude variation of these fluctuations is achieved through analysis of multiple confinement modes.
	 
	 The study presented here confirms longstanding concerns about broadening of microwave deposition\cite{gentle06}, and form a dataset which can be used to both drive and benchmark simulation efforts. This work considers only scattering by density fluctuations, although scattering by magnetic fluctuations is in principle possible\cite{Vahala}. Simulations of these discharges complement this experimental result, measuring a comparable degree of beam broadening on DIII-D\cite{thomas}. If ITER ECCD is broadened by the same degree measured here, constant power mode suppression requirements would increase by a factor of two or more. With modulation techniques which have already been explored on DIII-D\cite{koleman} and benchmarked by simulation\cite{refPolipreq}, the increase in power needed would be only 30\%. 	  

	Section 2 discuses a series of experiments with modulated microwave power performed with different edge conditions in an attempt to uncover a fluctuation broadening scaling law. Section 3 introduces a method of fitting heat transport to resolve deposition broadening in experimental data. Section 4 discusses the validation of this method against past transport studies.  An order of magnitude variation of the scattering fluctuations allows for a scaling of deposition broadening. Section 5 concludes the work with discussion of implications of microwave beam broadening for ITER.
	
	\section{Experimental Fluctuation Measurements in DIII-D }	
	\subsection{Measuring Fluctuations}
\begin{figure}
	\includegraphics[width=8cm]{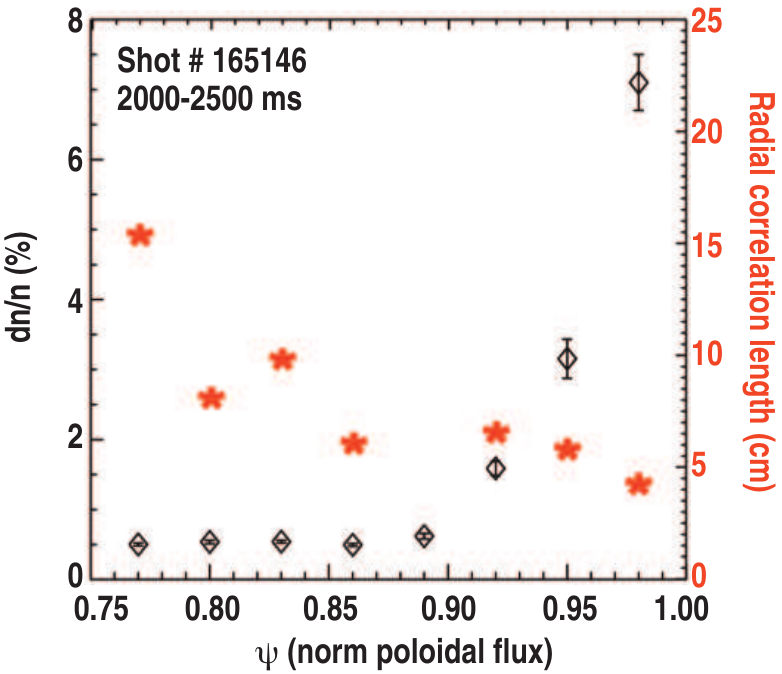}
	\caption{Density fluctuation measurements from beam emission spectroscopy made for a DIII-D H-mode broadening case. These larger scale fluctuations inform simulations made to understand the poloidal structure and correlation length of fluctuations.}
	\label{fig-BESnefluct}       % Give a unique label
\end{figure}

Average edge fluctuations in these discharges are measured by a combination of 2D beam emission spectroscopy (BES)\cite{mckee07} and Doppler Backscattering (DBS)\cite{refRhodbs}. BES measures larger scale density fluctuations through beam ion spectroscopy, and is sensitive to a larger scale, low poloidal wavenumber turbulence, with $k_\theta\rho_i<1 cm^{-1}$. The BES array located on the midplane captures an absolutely calibrated fluctuation amplitude and poloidal correlation length. An example of such profiles shown in Fig. \ref{fig-BESnefluct}. These measurements are used in the associated simulations\cite{thomas} to inform the spatial structure of the fluctuations.

\begin{figure}
	\includegraphics[width=8cm]{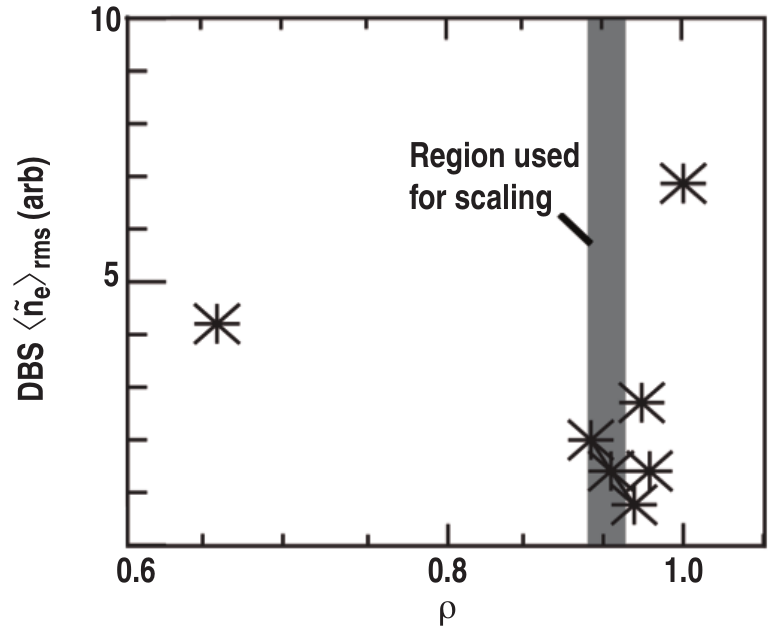}
	\caption{Density fluctuation measurements from Doppler backscattering made for a DIII-D H-mode broadening case. The amplitude of the fluctuations at $\rho=.95$ are taken as characterizing the turbulence amplitude. It is in this region where the turbulence drive term, the density gradient, is largest. Radial uncertainties are approximately $\Delta \rho = .05$. Thus the channel value at $\rho=.95$ can be considered an average over the entire steep gradient region.}
	\label{fig-DBSnefluct}       % Give a unique label
\end{figure}

Doppler backscattering measures smaller scale density turbulence, with a poloidal wavenumber wavenumber $1 cm^{-1}<k_\theta\rho_i<4 cm^{-1}$ \cite{refRhodbs}. The vacuum wavelength of the 110 Ghz microwave power is .36 cm, so scattering structures of this scale can be observed by the DBS system. DBS is cross-calibrated across these discharges, but is not absolutely calibrated to a $\tilde{n}_e$. However, the DBS fluctuation amplitude is the best available proxy for the density of edge structures on the millimeter scale, which were shown in previous simulations to scale strongly with deposition broadening\cite{refDec}. A time-averaged DBS amplitude from an H-mode shot is shown in Fig \ref{fig-DBSnefluct}.  

Edge structures move past the RF beam and turn over quickly $\mathcal{O}$(10 kHz) \cite{dudson16}, altering the deflection of the microwave beam about its equilibrium path.  This is much faster than the thermal confinement timescale $\tau_e \approx 100 ms$ which governs heating, and current relaxation time scale $\tau_r \approx 1s$ which governs current drive \cite{luce2001}. In these experiments, scattering is considered in terms of its time-average effect on the deposition profile. Simulations find that scattering is substantial only in the tokamak edge, producing an effectively broadened RF beam\cite{thomas}. The mean level of DBS measured millimeter scale density fluctuations at a normalized minor radius of $\rho=.95$ will be shown to scale strongly with the measured deposition width.
	
\subsection{Generating a Scaling in Fluctuation Amplitude}
		A range of discharge conditions with significantly different edge character can be achieved on DIII-D. A range of edge conditions were used as a means to produce an order of magnitude variation in fluctuation amplitude. 
		
		 L-mode is an operating scenario with no special enhancements to confinement\cite{wessonTokamaks}. A substantial level of fluctuations associated with the extended density gradient at the edge of the machine drives significant turbulent transport\cite{dudson16}. In these experiments, millimeter-scale density fluctuations are higher when the plasma is in a diverted L-mode configuration, as opposed to an inner wall limited configuration.
		  
		   The formation of a high confinement H-mode is known to be related to zonal flow suppression of turbulence in the tokamak edge\cite{rhodes2002}. As such, the time integrated edge fluctuation level measured in H-mode between prompt flows driven by edge localized modes is far lower than that measured in L-mode. QH-mode is a modification to H-mode wherein the pedestal is stabilized by a series of oscillations which have been found to lead to increased short wavelength turbulence\cite{rost}. 
		   
		   Negative triangularity (-$\delta$) L-mode discharges, a novel concept run on DIII-D, found the lowest levels of fluctuations. Similar discharges on TCV found substantially reduced turbulence was driven by changes in the shape-dependent trapped electron mode turbulence\cite{marinoni}.
		
			By considering the variation of scattering-relevant fluctuations across discharge conditions, an experimental scaling covering a substantial range in amplitude is produced. For example, in a shape-matched discharge the transition between L- and H-mode is associated with a factor of four drop in fluctuation amplitude measured by Doppler backscattering.	
			\subsection{Defining a Broadening Factor}
\begin{figure}
				\includegraphics[width=8cm]{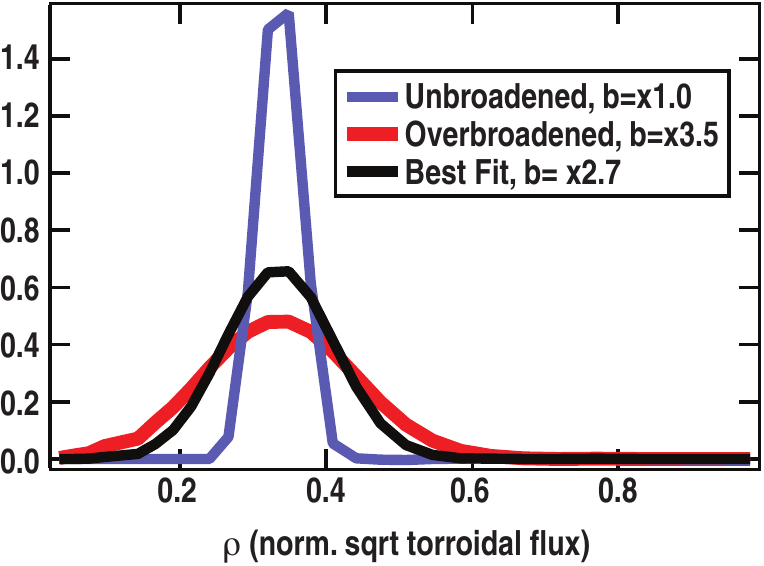}
				\caption{Power deposition functions for three different degrees of broadening in are shown. The selected diverted H-mode discharge used to evaluate these profiles shows the greatest degree of broadening of any study. In the next section that, it will be shown in that a fitting consistency with a transport model can resolve which of these profiles applyis correct.}
				\label{fig-Heat}       % Give a unique label
			\end{figure}

			Direct comparison of microwave deposition widths across discharge conditions is complicated by the temperature and density changes which define them. The TORAY-GA ray tracing code already accounts for the effects of geometry and temperature on propagation, but does not treat fluctuations\cite{refPrat}.  Power deposition width will scale linearly with beam width, thus the ratio of the best fit experimental width to the default TORAY-GA width is taken as the observed broadening factor. This ratio is defined as $b=FWHM_{Broadened}/FWHM_{TORAY}$, and will be used to quantify broadening in these experiments. For this work, a set of broadening Gaussians with a FWHM in $\Delta\rho$ ranging from .005 to .01 were applied to the dataset. This produces a b from 1 to at least 3.5, depending on the inherent deposition width. Inherent width is geometry dependent, but is generally of the order $\Delta\rho \approx.003$. A comparison of two broadened and an unbroadened heating profiles is shown in Fig. \ref{fig-Heat}.
			
	\section{Using Heat to Resolve Microwave Deposition}
	\subsection{Perturbation Measurements}
	Measurement of the microwave-driven heat perturbation in tokamaks is complicated by transport. For this work, gyrotron power is modulated from 10\% to 100\% of rated power with a 50\% duty cycle. A square wave modulation produces a regular heat pulse with the same harmonic content, allowing for frequency-domain analysis of the heat flux. In the absence of transport effects, the change in electron stored energy could be calculated as a function of density $n_e$, and temperature $T_e$ at the switch-on of ECH power\cite{refZer}. Perturbation measurements would then directly provide the power deposition as a function of radius.
	
	\begin{equation}
		3/2 n_e \frac{\partial T_e(\rho,t)}{ \partial t} \approx p_{ECH}(\rho,t)
	\end{equation}

\begin{figure}
	 	 	 	\includegraphics[width=.8\linewidth]{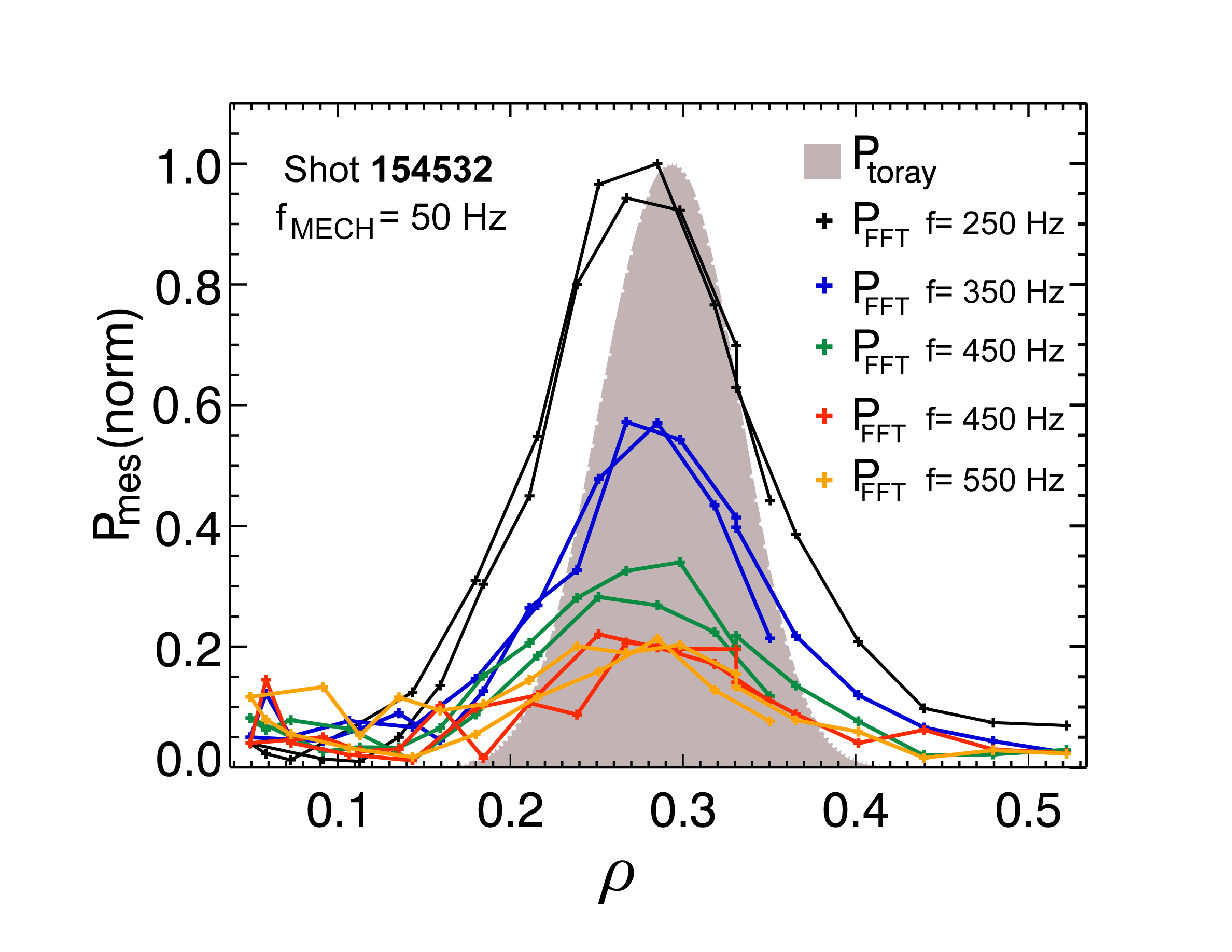}
	 	 	 	\caption{The Fourier analyzed temperature perturbation generated by modulated heating are shown. A square wave modulation of ECH power produces a harmonic set of perturbations, with higher harmonics reflecting faster timescale behavior. The 5th to 11th harmonics are shown, all substantially wider than the TORAY-GA source function in grey.}
	 	 	 	\label{fig-fTte}       % Give a unique label
	 	 	 \end{figure}

	 The perturbation from microwave heating on DIII-D is resolved by measurements of electron temperature ($T_e$). A 48 channel, absolutely calibrated, profile electron cyclotron emission (ECE) radiometer provides a 1D $T_e$ profile digitized at 500 kHz\cite{refAJece}. Second harmonic X-mode ECE coverage at $B_t= 2 \textmd{T}$ in DIII-D extends inwards from the optically thin scrape off layer at $\rho=1$, where $\rho$ is a radial coordinate normalized to square root of toroidal flux. Fourier analysis produces a temperature perturbation profile (modulated quantities are expressed with a tilde, for example $\tilde{T}_e$) which is compared with the ray tracing derived deposition profile in Fig. \ref{fig-fTte}. Experimental perturbation profiles are always wider than TORAY-GA calculated deposition.
	 
	 Transport fitting across harmonics can resolve the difference between deposition broadening, which is modulation frequency independent, and transport, which evolves over time. At least some of the difference between TORAY-GA and $\tilde{T}_e$ is due to transport. The perturbation applied to the plasma leads to a prompt change in the flow of energy, such that $\partial T_e/\partial t \ne dT_e/dt$. This produces a perturbation to the electron heat flux $Q_e$ which obscures the base deposition profile. 
	 
	 \subsection{Heat Flux Calculations}
	 The plasma response to modulated electron heating is dominated by a 1D temperature perturbation. The perturbation of electron density, plasma rotation, and turbulence for modulation frequencies above 10Hz are minimal, consistent with past experiments on DIII-D.\cite{gentle06, ernstiaea} The relationship between flux $Q_e$, stored energy, and power sources can be treated through energy conservation as represented by Eq. \ref{1dnrg}. 
	 
	 \begin{equation}
	 \label{1dnrg}
	 \frac{d}{dt}\left(\frac{3}{2}n_e \tilde{T}_e\right)- \nabla \cdot \tilde{Q}_e = \tilde{P}_{ECH}+\tilde{P}_{OTHER}
	 \end{equation}
	 
\begin{figure}
\centering
\includegraphics[width=0.7\linewidth]{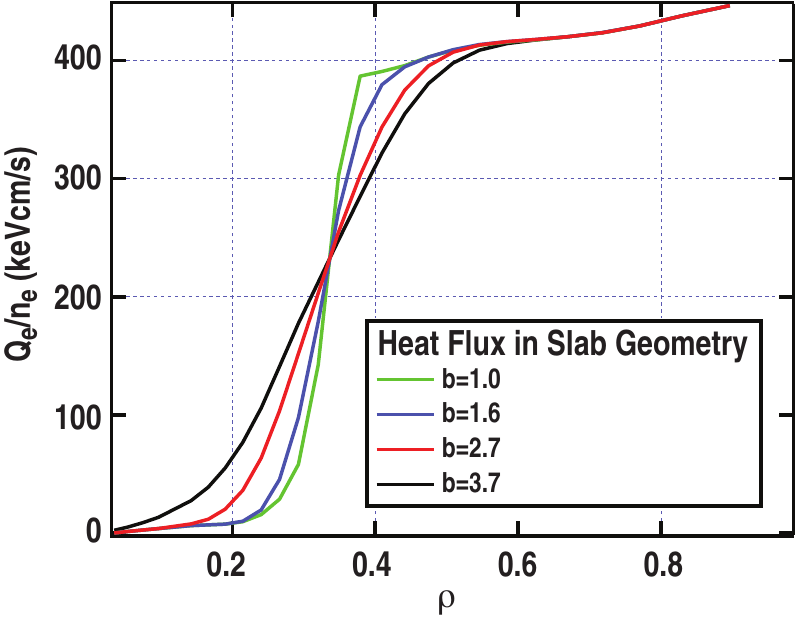}
\caption[Example heat fluxes w/ broadening]{The heat flux per particle is displayed vs $\rho$ for a set of broadening factors. In this discharge (a diverted L-mode), the perturbation is extended radially, so there are contributions far from the source region.}
\label{fig:heatfluxvsb(flat)}
\end{figure}

	 This work solves the heat equation for a set of heat fluxes through integration of the measured temperature perturbation and a trial $p_{ECH}$ profile ($\tilde{p}_{ECH}$ being the differential form of $\tilde{P}_{ECH}$). When treating the transport in cylindrical geometry, the heat flux contains an inherent $1/r$ radial variation, rather than the step character found for the same calculation in a slab geometry used in past studies\cite{refDeB}. 	Using separation of variables in r and t, the time-dependent portions can be Fourier transformed with the kernel $e^{2\pi i nf_{mod} t}$ into a harmonic series in modulation frequency, $f=f_{mod}*n$. Terms without a time dependence pass through as constants. 
	 
		\begin{equation}
		\tilde{Q}_e\left(r,f_{mod}*n\right)= \frac{1}{r} \int_{0}^{r}\left( \tilde{p}_{ECH}(r',f_{mod}*n) - 3i\pi nf_{mod} n_e(r') \tilde{T}_e(r',f_{mod}*n) \right)r' dr
		\label{heatfluxcyl}
		\end{equation}       
	For this analysis, a Gaussian broadening factor is applied to $\tilde{p}_{ECH}(\rho)$. The degree of broadening is quantified, as defined earlier, by a ratio of broadened and unbroadened profile widths, $b$. An example of the different heat fluxes that result from this broadening are shown in Fig \ref{fig:heatfluxvsb(flat)} . The consistency of the fluxes in the edge, far from the deposition region, gives confidence that the power is being conserved properly by the Gaussian blur.
	
	\subsection{Defining Transport Coefficients}
	\label{TransportFit}
	The consistency between the calculated 1D heat flux and a 1D transport model is used in this work as a metric to evaluate broadened power deposition.  Strictly diffusive models fail to capture heat which flows against the gradient in many microwave heated plasmas\cite{refMan}. Following the form of Ryter \textit{et al}\cite{refRyter}, diffusive (denoted by the coefficient D) and convective (denoted by the coefficient V) components are necessary to capture the effects of microwave heating. Expressing this as an equation for heat flux, $Q_e$:
	
	 \begin{equation}
	 Q_e\left(\rho, t\right)=-D n_e \nabla T + Vn_eT.
	 \label{key}
	 \end{equation}

	The D and V coefficients are not necessarily constant. They can vary both radially and over the power modulation cycle, with an unknown dependence on tokamak parameters. Linearizing this form in time allows modulation-independent background heat flux to be seperated from the modulation induced portion, denoted $\tilde{Q}_e$.
	\begin{equation}
	\tilde{Q}_e= n_e(-D\nabla\tilde{T}_e-\tilde{D}\nabla T_e+ V\tilde{T}_e+\tilde{V} T_e ) + \tilde{n}_e (-D \nabla T_e + VT_e) 
	\label{qexpand}
	\end{equation}
	
	Rather than expanding directly in time, $\tilde{D}$ and $\tilde{V}$ are expressed as a set of partial derivatives in the dominant $\tilde{T}_e$ modulation\cite{refGen88}.
	
	\begin{equation}
	\tilde{D}=\nabla\tilde{T_e} \frac{\partial D}{\partial \nabla T_e}+ \tilde{T_e} \frac{\partial D}{\partial T_e}+ \tilde{n}_e \frac{\partial D}{\partial n_e } +\nabla \tilde{n}_e \frac{\partial \nabla D}{\partial \nabla n_e }
	\label{Dmoddep}
	\end{equation}
	
		\begin{equation}
		\tilde{V}=\tilde{T}_e\frac{\partial V}{\partial T_e}+\nabla\tilde{n}_e \frac{\partial V}{\partial \nabla n_e}+\tilde{n}_e \frac{\partial V}{\partial n_e}
		\label{Vmoddep}
		\end{equation}

	\subsection{Grouping Modulated Terms into a Fitting Equation}
	The heat flux expressed by the transport model is complicated, Eq \ref{qexpand} has a number of dependencies. Inserting equations \ref{Dmoddep} and \ref{Vmoddep} into the flux formulation Eq \ref{qexpand} produces an equation for modulated heat flux. The dependences in these terms are complex, but we can use past work\cite{refGen88} to guide our grouping into a set of independent fitting parameters based on the dependence on the temperature perturbation $\tilde{T}_e$. A term proportional to $\tilde{T}_e$, to $\nabla \tilde{T}_e$, and one which depends on neither are used. These can be understood respectively as a convective, diffusive, and coupled heat flux. Sorting of this form is helpful, as temperature perturbation is much larger than the density perturbation, which is seen to be less than 1\% in coherently averaged interferometer data.

	\begin{multline}
	\label{biglin}
	\tilde{Q}_e= -\nabla \tilde{T}_e (D+\nabla\ T_e \frac{\partial D}{\partial \nabla T_e} )+ \tilde{T}_e (V   + T_e \frac{\partial V}{\partial T_e}-\nabla\ T_e \frac{\partial D}{\partial T_e}) \\ +\nabla \tilde{n}_e (T_e\frac{\partial V}{ \nabla n_e}- \nabla T_e \frac{\partial D}{ \nabla n_e})+ \tilde{n}_e (T_e(\frac{V}{n_e}+\frac{\partial V}{\partial n_e}) -\nabla T_e (\frac{D}{n_e}+\frac{\partial D}{\partial n_e}))
	\end{multline}

	The modulated diffusion term can be assembled from all terms proportional to $\nabla \tilde{T}_e$:
	\begin{equation}
		D_M=D+\nabla\ T_e \frac{\partial D}{\partial \nabla T_e} 
	\end{equation}
		
	Similarly a modulation convection can be assembled from all terms containing $ \tilde{T}_e$
	\begin{equation}
	V_M=V   + T_e \frac{\partial V}{\partial T_e}-\nabla\ T_e \frac{\partial D}{\partial T_e}
	\end{equation}
	
	The remaining terms are combined into coupled transport, $\tilde{\xi}$. 
	
\subsection{Coupled Transport}	  
	 Coupled transport, $\tilde{\xi}$, contains terms driven by the modulated density and its gradient and any other small corrections. These could include an apparent flux from motion of ECE measurement locations due to plasma shift, non-thermal transport, modulated ion-electron exchange power, and Ohmic power modulation. These are grouped together as  $\tilde{\xi}$ for the fit as they are functionally independent of the perturbed terms.
	 \begin{equation}
	 \label{xidef} 
	 \tilde{\xi}=
	  \frac{\nabla\tilde{n}_e}{n_e}(T_e\frac{\partial V}{ \nabla n_e}- \nabla T_e \frac{\partial D}{ \nabla n_e})+ \frac{\tilde{n}_e}{n_e} (T_e(\frac{V}{n_e}+\frac{\partial V}{\partial n_e}) -\nabla T_e (\frac{D}{n_e}+\frac{\partial D}{\partial n_e}))
	 \end{equation}
	 
	  This $\tilde{\xi}$ term must be given some frequency dependence to simultaneously fit fluxes across harmonics. The physical interpretation of this fact is that the $\tilde{\xi}$ term must evolve over the modulation cycle much as the other terms do. As used in past ballistic heat pulse work by Fredrickson\cite{Fredrickson}, an exponential decay with a free phase can capture the dominant time response of the coupled transport. This term could also in principle capture the effects of fast electron transport modification to diffusion. The exponential decay of the form $f(t)=A*exp(-ct)$ has a Fourier transform $F(f)=2A/(c+i2\pi f)$. The modulation of diffusion is not necessarily in phase with the turn on of the ECH perturbation, or the peak of temperature. Pulling the imaginary component out as a phase term $\theta_\xi$, and the amplitude as $\xi_0 = A\sqrt{c^2+(2\pi f)^2}$ allows simplification to the fitting form of $\tilde{\xi}$:
	  
	 \begin{equation}
	 	\tilde{\xi}(r,f)=\xi_0(r)\times e^{i\theta_\xi} \times  \frac{4\pi c}{4\pi f_{mod}^2+c^2}
	 \end{equation}

 Writing a fit term with this freedom allows a ballistic population without enforcing it. Both diffusion and convection are present to account for bulk components of the heat flux. Estimates for ITER parameters suggest that the level of fast electron transport predicted $(D_{rr}\approx .15 m^2s^{-1})$ will not lead to a substantial broadening of the ECCD profile\cite{casson}. DIII-D studies which did not treat fluctuation broadening capped the levels of diffusive transport for fast electrons at a minimal $D_{rr}=0.4 m^2s^{-1}$\cite{pettyfst}. While transport at this level can drive profile broadening, a value of $D_{rr}=1.0 m^2s^{-1}$ is considered the minimum level for deleterious broadening\cite{casson}. The observed broadening for L-mode is consistent to within uncertainties for both core and edge deposition in DIII-D, whereas the work of Harvey \textit{et. al.}\cite{harvey} predicts two orders of magnitude difference between edge and core current diffusion times. If diffusion of hot particles confounded broadening, $b$ factors for core and edge deposition would differ, which is not found to be the case. Thus this work considered fast electron effects as being addressed by the $\tilde{\xi}$ term.
 
 Fitting finds the amplitude of the coupled transport driven flux to be smaller than either the diffusive or coupled terms. Only a fraction of the flux (15\% at most) comes from coupled transport in the fitting performed in the next section. Thus this work does not attempt to reduce $\tilde{\xi}$ to its potential dependencies.
 	 
\subsection{The Reduced Transport Model}
These simplifications allow the complexity of Eq. \ref{biglin} to be reduced to a simpler parameter set which can capture the bulk profile dependencies of modulated heat transport. Writing the coefficients in their simplified form defined in the previous sections gives Eq. \ref{eqn:lintranmod}.
\begin{equation}
\tilde{Q}_e(r,t)/n_e(r,t)=- D_M \nabla \tilde{T}_e+V_M \tilde{T}_e+\tilde{\xi}
\label{eqn:lintranmod}
\end{equation}

The consistency of this transport model with a chosen deposition function can be evaluated.  Eq. \ref{eqn:lintranmod} is compared to the heat flux from energy conservation, Eq. \ref{heatfluxcyl} and agreement between these two forms of flux is used as a check on ECH deposition, as the energy conservation derived heat flux is sensitive to $\tilde{p}_{ECH}(r,f)$.

\section{Transport Fits to Broadened Deposition}
\subsection{Methodology}
\begin{figure}
	 	\includegraphics[width=8cm]{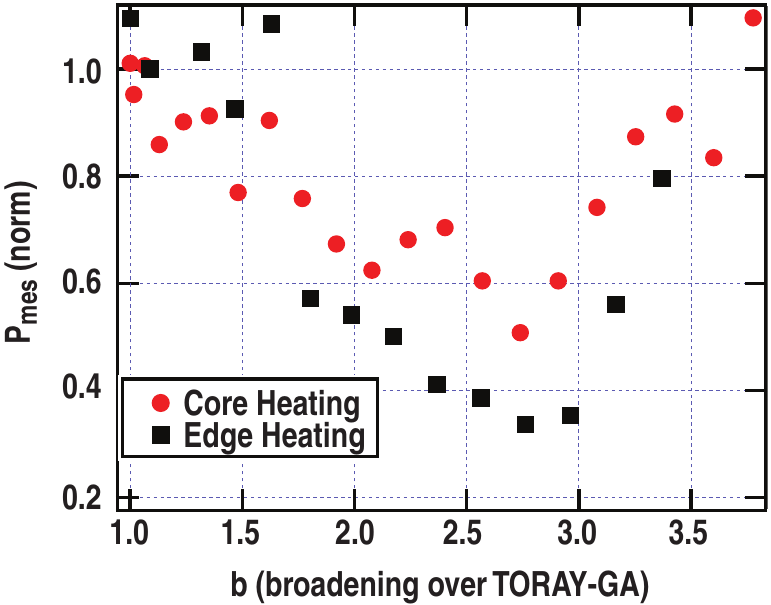}
	 	\caption{$\chi^2$ minimization is used as a means of evaluating the choice of broadening factor, b. When depositing at near the core $(\rho=.25)$ and near the edge $(\rho=.6)$ in same plasma, inherent widths of deposition differ sharply due to changes in plasma parameters and geometry with increasing radius. The fluctuation amplitude in the edge is matched, however, and the same broadening factor produces the best fit in both cases. This gives confidence in the use of b as a simple parameter for scaling.}
	 	\label{fig-chiSQn}       % Give a unique label
	 \end{figure}

 Broadening of the microwave deposition profile can be performed using a variety of transformations; a power conserving Gaussian blur is used for this work to alter the power deposition used to calculate flux in \ref{heatfluxcyl}. This is set equal to Eq. \ref{eqn:lintranmod} to form Eq. \ref{eqAG} and fit using propagation of uncertainties in each parameter. 
	\begin{equation}
		 \label{eqAG}
		 - D_M \nabla \tilde{T}_e+V_M \tilde{T}_e+\tilde{\xi}= \frac{1}{rn_e} \int_{r'=0}^{r}( \tilde{\rho}_{ECH}(r',f) - 3i\pi f n_e(r') \tilde{T}_e(r',f) )r' dr
	 \end{equation}
	 The goodness of fit parameter, $\chi^2$ shows a clear minimum with increasing broadening for fits under a range of discharge conditions. Figure \ref{fig-chiSQn} shows an example of the broadening factors explored for an L-mode shot which experienced both core and edge $\tilde{p}_{ECH}$ modulation.  While injection angle and deposition region change the TORAY-GA widths ($\Delta\rho_{core}=.275$, $ \Delta\rho_{edge}=.233$), the minimization for both cases occurs for factor of $b \approx 2.7$.  All discharges studied have a similar minimization, although broadening factors observed vary from $\times 1.4 - \times 2.8$ . Broadening to a degree which minimizes $\chi^2$ is also found to produce consistently positive modulated diffusion coefficients.
	 	 	 	 	  
	A constant broadening factor is found for core and edge deposition cases, even though fast transport differs dramatically between tokamak core and edge\cite{harvey}. 
\subsection{Example Transport Coefficients}
\begin{figure}
		\includegraphics[width=8cm]{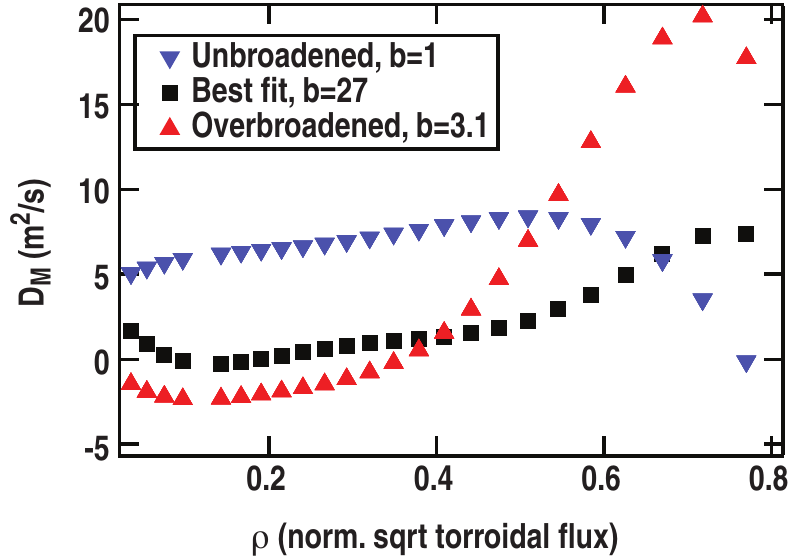}
		\caption{In this case, $D_M$ is fit with polynomial as a function of $\rho$. $D_M$ is strongly sensitive to the phase of the heat flux, as the gradient perturbation $\nabla \tilde{T}_e$ flips phase across the deposition region. While $D_M$ can in principle be negative, this would indicate transport barriers or other phenomenon driving a negative dependence of flux on temperature gradient. Radially increasing, consistently positive values of $D_M$ for a significantly broadened L-mode coincide with minimization of $\chi^2$.}
		\label{fig-bplot}       % Give a unique label
	\end{figure}

	Transport coefficients in tokamaks such as DIII-D\cite{gentle06} and TFTR\cite{Fredrickson} were found to vary by an order of magnitude over the plasma radius.  Fitting can be performed with fixed coefficients over a small range, or through fitting a set of orthogonal polynomials. From Fig. \ref{fig-chiSQn}, it is found consistency between linear transport and energy conservation is maximized for a $~\times2.7$ broader FWHM for $P_{ECH}(r)$ compared with TORAY-GA. An example of a radial fit of $D_M$ in the previous L-mode discharge for various choices of $b$ is shown in Figure \ref{fig-bplot}. Polynomial fit results for the diffusion coefficient corresponding to that broadening are reasonable - consistently positive with no sharp negative excursions or peaks far from the heating region in L-mode cases where the best-fit broadening applies. This rule holds across the discharges studied. Additional fit results using locally constant coefficients are presented in the next section.

	\subsection{Comparison of Coefficients with the Differential Heat Pulse Formulation}
	 
	 This section demonstrates that the integral method can reproduce well the results of past transport studies. Validation of this heat flux-based fitting method has been performed against a set of modulated ECH data in L-mode. Constant-power gyrotrons were used to alter the temperature gradient at the plasma edge. Power was incrementally moved from edge to mid-radius to steepen the gradient, producing the 'critical gradient' behavior, where transport coefficents are found to be a function of temperature gradient above a certain value of gradient or in this case scale length $L_{T_e}= (\partial T_e/\partial r )/ T_e$\cite{refDeB}. Results from this study, based on the differential form of the heat equation, can be used as a benchmark for the integral heat flux method presented here. DeBoo solves the heat equation in a differential form, reproduced from his paper\cite{refDeB} as Eq. \ref{eqn:DeBoo}.
	 
	   The differential form of deBoo \textit{et. al.} has a modulated diffusion $D_{HP}$, convection $V_{HP}$, and a third damping term $1/\tau$, similar to $\xi$\cite{refDeB}.  In the differential form, a second derivative of $\tilde{T}_e$ appears instead of an integral.  
	
			 	\begin{equation}
		 	 -D_{HP}\nabla^2 \tilde{T}_e+V_{HP} \nabla \tilde{T}_e+ \tilde{T}_e (\frac{3\pi}f i  +\frac{1}{\tau}) = \tilde{P}_{ECH}/n_e
		 	 \label{eqn:DeBoo}
		 	 \end{equation}

\begin{figure}
	 	\includegraphics[width=8cm]{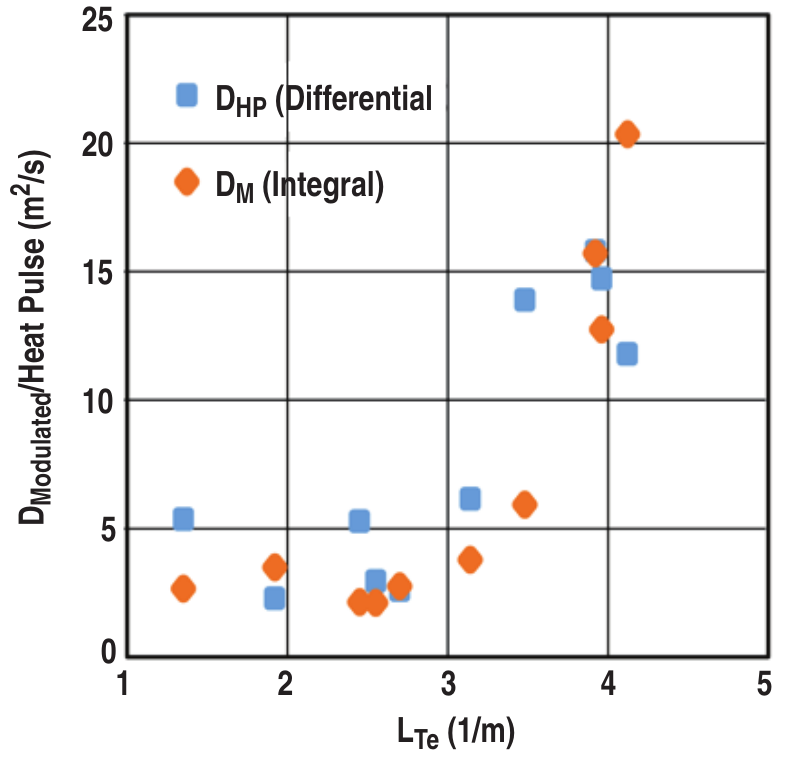}
	 	\caption{Integral method transport fitting can reproduce the critical gradient results from the perturbative experiment of deBoo \textit{et. al.}\cite{refDeB}. The modulated diffusivity measured at $\rho=.6$ increases sharply as the electron temperature scale length, $L_{T_e}$ approaches a critical value, here found to be $\approx 2.5 m^{-1}$. The differential and flux integral method fits of diffusion coefficent are evaluated in a slab geometry from $\rho=.45$ to $\rho=.75$}
	 	\label{fig-DmodComp}       % Give a unique label
	 \end{figure}

	   For experimental data, a second derivative is highly model dependent as are its uncertainties\cite{refZer}. For a Monte Carlo integral, error bars are directly calculated from the standard deviation of the random draw. Minimizing intrinsic uncertainty in the fit equation improves the method's sensitivity to changes in power deposition. However, in a slab geometry, the integral and differential form coefficients are directly comparable. Geometric factors do not enter the integral, and coefficients have no radial dependence.  Thus $\int D_{HP} \nabla^2\tilde{T}_e dx$ is equivalent to $D_M \nabla\tilde{T}_e$, with a similar relation for $V_M$ and $V$. To compare the two methods, a constant coefficient fit from $\rho=.2$ to $\rho=.6$ was made with both methods. The results for the diffusion coefficient are given in Fig. \ref{fig-DmodComp}. It is found that the integral method reproduces the same critical gradient behavior observed by the differential method.
	
		\subsection{Identifying a Scaling with Fluctuation Level}
		
		A range of deposition locations and edge conditions have been examined to produce a beam broadening scaling with fluctuation amplitude in the edge ($\rho\approx .95$). Across the dozens of discharges in the dataset, propagation angle and deposition location may not always be identical. Previous work has shown that fluctuation amplitude and path length through fluctuations define a broadening factor which is otherwise path independent. To produce a large variation in turbulence amplitude, intervals in a range of discharge conditions are considered. Selected intervals have in excess of 15 modulation periods, and show density variations less than $1\%$.
		
\begin{figure}
			\includegraphics[width=8cm]{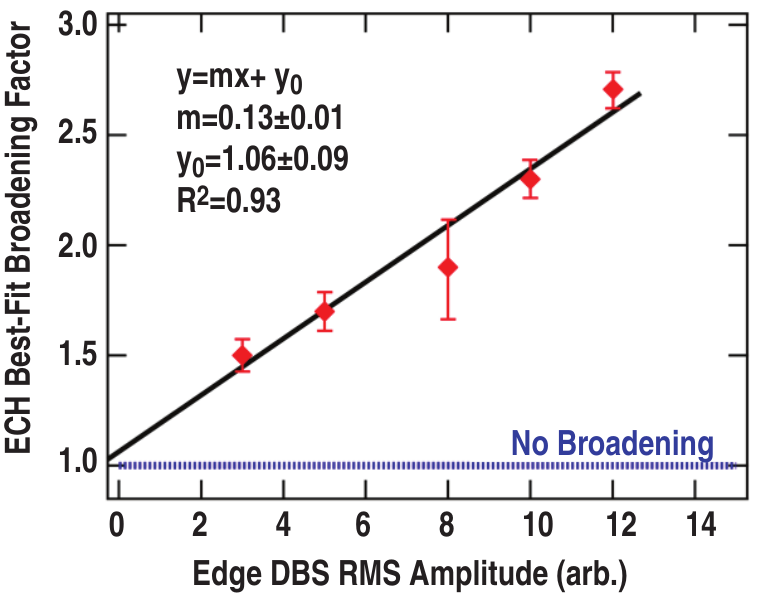}
			\caption{A collection of 19 discharges in 5 conditions have both fluctuation data and modulation needed to form a scaling. Broadening of the narrow RF beam used in deposition calculated in TORAY-GA will lead to a corresponding increase in deposition width. The horizontal axis is the normalized density fluctuation amplitude from the Doppler backscattering system at $\rho=.95$. The vertical axis is the factor increase in deposition profile width over TORAY-GA's equilibrium, no-fluctuation deposition FWHM in $\rho$.}
			\label{fig-bscale}       % Give a unique label
			\end{figure}

		Fit ranges are selected to avoid magnetic islands, and the ELM-perturbed edge of H-mode plasmas, which can extend in as far as $\rho=.6$\cite{leonardelm}. Figure \ref{fig-bscale} shows broadening factors derived from these fits assembled by discharge condition. Uncertainties in the experimental data are derived from the statistical interpretation of $\chi^2$. An increase in the reduced goodness of fit, $\chi^2_\nu$ , by a standard deviation is taken as an approximate $1\sigma$ value for these measurements, producing a 10-20\% uncertainty in broadening factor.
		
		The experimental scaling of broadening with DBS-derived fluctuations is linear. A linear scaling was predicted by J. Decker with the LUKE/C3PO code\cite{refDec}. In an accompanying paper submitted to this journal, a 3D full-wave cold plasma finite difference time domain code EMIT-3D\cite{williams14} has been used to simulate the extent to which scattering broadens a microwave beam in DIII-D\cite{thomas}. A full wave treatment is necessary when the inhomogeneity scale length is comparable to the wavelength \cite{koehn16}, explaining why this broadening is greater than found by past simulation efforts\cite{refPrat}. Results from 3D full wave simulations are consistent with the experimental broadening factors found in experiments.
	
	\section{Implications for ITER}
		The increased power requirements for power deposition broadening observed for H-mode on DIII-D would be tolerable on ITER. Broadening factors in the range $b=1-2$ have been simulated for current drive on ITER\cite{refPolipreq}. On ITER, both 3/2 and 2/1 tearing modes will be targeted for ECCD suppression. A total of 25 MW of EC power are planned for all purposes in ITER. The 3/2 mode is expected to require $\approx 7\space MW$ of constant power to suppress. Growth of this mode is slow, on the order 5 seconds, so gyrotrons will likely have time to turn on and track the island structure\cite{poliaps}. The 2/1 mode has a substantially higher risk of locking, and it can grow from seed islands with a width of 1 cm to a width of 5 cm in only 2-3 s\cite{poliaps}, at which point it risks locking and disrupting the plasma\cite{lahayeNF}.
		
		Failure to suppress these islands will limit the maximum Q achievable on ITER, even in the absence of locking. Unsuppressed islands cause a profile flattening, which can be estimated according to the formulation of Sauter and Zohm\cite{sauterpower}. An unsuppressed 2/1 mode will limit Q to 4, due to $T_i$ flattening, while an unsuppressed 3/2 limits Q to 7. However, it must be noted that fusion gain is impacted by both unsuppressed islands and by always-on EC power use. ITER is expected to use 50 MW of power in its baseline configuration, and thus with an expected fusion power of 500 MW, a gain of Q=10 can be achieved \cite{gjackson}. Constant use of the 25 MW of gyrotron power limits maximum gain to Q=6.7.
		
		Density fluctuation-driven broadening of ECH beams of the magnitude observed on DIII-D is plausible for ITER. While fluctuation levels are lower, ~30\% on DIII-D\cite{shafer12} as compared to ~10\% on ITER\cite{koehn16}, the path length through the plasma is over $\times3$ longer. Studies made using the beam-tracing WKBEAM\cite{guidi} code provide information on the power needed to offset deposition broadening on ITER\cite{refPolipreq}. The width of the beam relative to the current drive profile has been explored as a free variable.
		
		 When the width of the ECCD profile is less than that of the island, only the total current deposited is relevant. When power deposition is wider than the targeted island that suppression becomes more expensive as power is lost outside the island chain. With this in mind, estimates for a factor of $\times2$ deposition broadening, which is typical of these discharges, render full suppression with always-on power far more expensive, with a power requirement approaching the maximum achievable from a single launcher module, 13 MW\cite{refPolipreq}. Assuming use of narrower deposition from the lower launcher module, and correct alignment, power requirements can be estimated from the work of Poli \textit{et. al.} \cite{refPolipreq}. 3/2 suppression requirements will nearly double, to 12 MW. Power requirements for the wider 2/1 mode are increased by a smaller factor, to 10 MW. Factors observed on DIII-D for L-mode, of the order $b=2.5$, would require techniques such as power modulation\cite{Kasparek} to suppress the 2/1 mode with the power from a single launcher. These requirements assume a similar magnitude of broadening will be observed on ITER, as on DIII-D. A physics-based projection using experimentally-benchmarked simulation codes, capable of resolving the effects of broadening, is needed to quantify power requirements.
	
	\section{Conclusions}
	\label{Conclusions}	
	In this work, a linear transport model was used to fit the electron heat flux generated by electron cyclotron heating. Transport fit with a 3-term model which includes diffusion, convection, and coupled transport effects is optimized for a broadened deposition, which scales linearly with density fluctuation amplitude across a range of discharge conditions. The simulation paper accompanying this work finds a consistent degree of broadening in 3 DIII-D cases where full wave analysis has been performed. Experimental broadening factors are found to be consistent with full wave simulations based on these experiments\cite{thomas}. Beam width was found to be x1.7-2.8 wider than predicted by ray tracing through a no-fluctuation equilibrium, increasing linearly with edge turbulence amplitude in both cases. Experiments backed by theory suggest a significant increase in microwave deposition widths on DIII-D. This leads to an increase in power requirements for mode control. Results from these experiments can be used in future work to produce physically benchmarked simulations of mode control power requirements on ITER.

\small{ This material is based upon work supported by the U.S. Department of Energy, Office of Science, Office of Fusion Energy Sciences, using the DIII-D National Fusion Facility, a DOE Office of Science user facility, under Award DE-FC02-04ER54698 and DE-FG03-97ER54415. DIII-D data shown in this paper can be obtained in digital format by following the links at \url{https://fusion.gat.com/global/D3D\_DMP}}

%%%%% CLEAR DOUBLE PAGE!
\newpage{\pagestyle{empty}\cleardoublepage}

\end{document}